# THE EFFECTS OF ADDITIVES ON THE PHYSICAL PROPERTIES OF ELECTROFORMED NICKEL AND ON THE STRETCH OF PHOTOELECTROFORMED NICKEL COMPONENTS

D.M.Allen[1], N.Duclos[1], I.Garbutt[2], M.Saumer[3], Ch.Dhum[3], M.Schmitt[3] and J.E.Hoffmann[3]

[1]School of Industrial and Manufacturing Science, Cranfield University,
Bedford MK43 0AL, UK
[2]Tecan Ltd, Granby Industrial Estate, Weymouth, Dorset DT4 9TU, UK
[3] University of Applied Sciences Kaiserslautern, D-66482 Zweibrücken
and D-67657 Kaiserslautern, Germany

**ABSTRACT**

The process of nickel electroforming is becoming increasingly important in the manufacture of MST products, as it has the potential to replicate complex geometries with extremely high fidelity. Electroforming of nickel uses multi-component electrolyte formulations in order to maximise desirable product properties. In addition to nickel sulphamate (the major electrolyte component), formulation additives can also comprise nickel chloride (to increase nickel anode dissolution), sulphamic acid (to control pH), boric acid (to act as a pH buffer), hardening/levelling agents (to increase deposit hardness and lustre) and wetting agents (to aid surface wetting and thus prevent gas bubbles and void formation). This paper investigates the effects of some of these variables on internal stress and stretch as a function of applied current density.

## 1. INTRODUCTION

The process of nickel electroforming is becoming increasingly important in the manufacture of MST products, as it has the potential to replicate complex geometries with extremely high fidelity. In order to reduce the costs per piece of polymeric components, replication from electroformed precision moulds by micro-injection moulding or hot-embossing is attracting the attention of manufacturers of microfluidic devices and diagnostic/analysis micro-systems as it appears to be one of the few economically-viable processes available to industry when product demand exceeds millions of components per annum [1].

Nickel is the most commonly-used electroformed material in such applications as it may be deposited with minimal internal (residual) stress. It is a primary requirement that the electroformed metal replicates the mould geometry and dimensions precisely and does not distort after separation from the mould mandrel. However, this is not easily done, requiring strict temperature and humidity control especially if the mould is made by imaging and developing photoresist; a process known as photoelectroforming (PEF). A well-known phenomenon in PEF is the inadvertent production of parts where an expansion of dimensions occurs after release from the mould, commonly known as "stretch".

Electroforming of nickel uses multi-component electrolyte formulations in order to maximise desirable product properties. In addition to nickel sulphamate (the major electrolyte component), formulation additives can also comprise nickel chloride (to increase nickel anode dissolution), sulphamic acid (to control pH), boric acid (to act as a pH buffer), hardening/levelling agents (to increase deposit hardness and lustre) and wetting agents (to aid surface wetting and thus prevent gas bubbles and void formation).

Several distinct types of nickel can therefore be deposited from such a range of nickel electroplating compositions, including:
- Soft nickel, based on a nickel sulphamate electrolyte
- Hard nickel, based on a nickel sulphamate electrolyte with an additional proprietary hardener and
- Ni-speed, a high-speed, high-concentration, nickel sulphamate formulation, developed originally by International Nickel.

The quantity of metal deposited follows Faraday's Laws of Electrolysis and is proportional to the applied current and time. To reduce deposition time, it is therefore advantageous to electro-deposit with a high current. Nickel electroforms can thus be fabricated to provide a range of characteristic properties including variable grain size, Young's modulus, yield strength, ultimate tensile strength, hardness and wear rate over a wide range of current densities [2].

## 2. STRESS THEORY

Residual stress is a tension or compression that exists in the bulk of a material without application of an external load. It can exist in a macro form over a few grains ($\sigma_I$), over one particular grain ($\sigma_{II}$), or even within a grain across several interatomic distances ($\sigma_{III}$). The last two stresses may be termed microstresses [3] [4].

In electroformed metals such as nickel, the stress may be either tensile or compressive. In the case of tensile (+) stress, the deposit can be likened to an expanded spring that is trying to contract relative to the substrate (mandrel) that it is attached to. In the case of compressive (-) stress, the deposit, likened to a compressed spring, is trying to expand relative to its mandrel [5].

As mentioned in the Introduction, electroforming of nickel uses multi-component electrolyte formulations and each component will affect the degree of stress in the deposit produced. In addition, internal stress will be affected by current density, temperature, agitation, plating cell geometry, the composition and condition of the anodes, the anode/cathode surface area ratio, the quality of the DC power supplied and the nature of the mandrel cathode. Indeed, it is hard to find a process variable that does not influence the deposit internal stress as pointed out by Stein [5].

In industrial electroforming and PEF, the tolerances of the parts are often very tight and accurate dimensions are difficult to achieve even in a rigorous temperature-controlled environment. In considering the release of a compressively-stressed deposit from its mandrel, the electroformed part will expand and change its dimensions such that the part may be out of tolerance and a costly reject. It is therefore vital that stress is minimized towards zero in order to obtain separated parts within dimensional tolerance. The stress cannot be controlled unless it can be measured accurately. Two methods for stress measurement have therefore been evaluated so that the results may be compared.

## 3. STRESS MEASUREMENT METHODS

### 3.1 Deposit stress analyzer

The deposit stress analyzer (Figure 1) is a popular instrument for measuring stress in a production environment as it offers significant cost- and time-savings in comparison with other methods [5, 6].

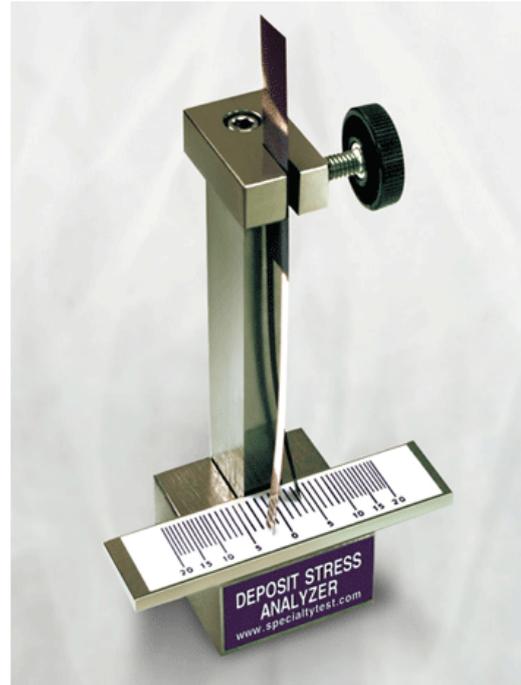

Figure 1. Deposit stress analyzer

The method comprises fixing a strip of metal, slit into two "legs" that are coated on opposite sides to each other with an insulating lacquer, in a special test cell within the plating tank. As electroplating of the mandrel is carried out, metal also deposits on opposite sides of the metal legs. With the production of stressed deposits, the legs bend and the resulting separation between them can be measured directly as shown in Figure 1. As the stress in the deposit increases, the separation will become greater. A calibration chart allows conversion from the separation distance to the internal stress value.

### 3.2 X-ray diffraction (XRD)

The strain ($\varepsilon$) in the direction normal to the diffraction planes is related to the corresponding lattice spacing (d) that is linked directly to the diffraction angle ($\theta$) by Bragg's Law: $n\lambda = 2d.\sin\theta$. For each inclination of the sample, defined by two angles, $\psi$ and $\varphi$, the strain $\varepsilon$ is thus related to the corresponding lattice spacing d.

The relationship between the strain $\varepsilon$ and the stress tensor components $\sigma$ is then derived, under specific conditions, through the $\sin^2\psi$ law [7] such that:

$\varepsilon = \frac{1}{2}S_{2\{hkl\}}\sigma_\varphi \sin^2\psi + s_1(\sigma_1 + \sigma_2)$

Thus a plot of $\varepsilon$ *vs*. $\sin^2\psi$ gives a straight line, the slope of which allows the calculation of the stress component $\sigma_\varphi$, in the direction $\varphi$ of the surface if the elastic constant ($\frac{1}{2}S_{2\{hkl\}}$) is known [3].

The diffractometer used to measure the stress in the electrodeposited material is shown in Figure 2.

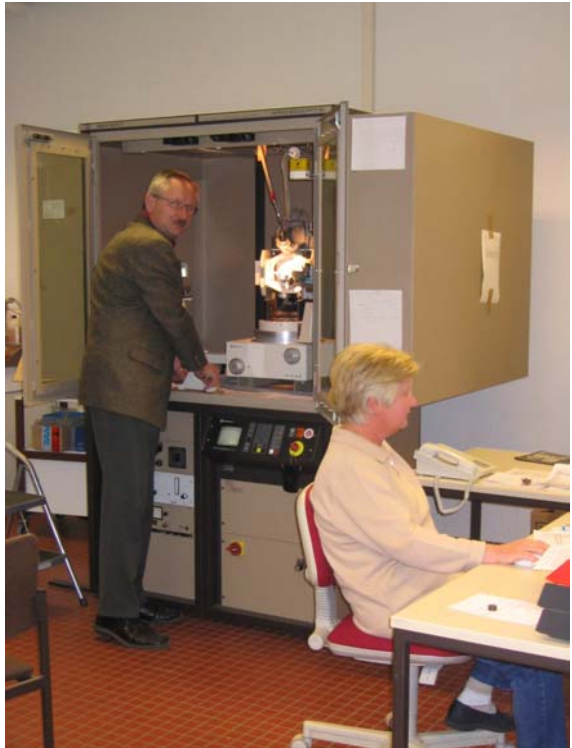

Figure 2. Seifert ψ-Diffractometer XRD 3000 PTS

**3.3 Comparisons between the two methods of measuring stress**

There are many differences between the two aforementioned methods used for stress measurement. The small-area deposit stress test strip is quick to use (<20 minutes) but can never be positioned in the exact location of the cathode and may not receive the average cell current density as it can vary considerably across a wide-area mandrel. The test strip is also orientated perpendicular to the mandrel to prevent shielding of one side of the strip. This results in a macro-measurement that averages the overall effects of the different current densities and micro-stresses in the deposit.

On the other hand, the XRD equipment only measures to an average penetration depth of 3.5μm over a very small area of 0.126mm$^2$. The analysis time is long and takes 3.5 hours.

## 4. RESULTS OF STRESS MEASUREMENTS

**4.1 Samples**
Nickel deposits, approximately 20μm thick, were electroplated onto large brass mandrels (48.8cm x 80.4cm) at current densities of ~1Adm$^{-2}$ and ~5Adm$^{-2}$.

Smaller samples (2cm x 5cm) were cut out from the middle of these plated mandrels and used for the XRD investigations, firstly whilst still attached to the brass mandrel and secondly when manually stripped from the mandrel. These samples were compared with the results obtained from the deposit stress analyzer samples electroplated under the same electrolytic and temperature conditions. The results obtained are shown in Table 1.

| Sample and current density | Internal Stress (MPa) | | |
|---|---|---|---|
| | (On brass) from deposit stress analyzer | (On brass) from XRD | (Free-standing) from XRD |
| St 56 ~1Adm$^{-2}$ | - 21 ± 10 | - 50 to - 100 ± 15 | - 37 ±13 |
| St 57 ~5Adm$^{-2}$ | + 1± 10 | - 120 ± 20 | - 21 ±50 |

Table 1. Results of stress analysis by the two methods

The results for the XRD analyses are plotted in Figure 3 and show some variation according to the angle of measurement and general movement towards zero internal stress, when released from the mandrel.

The highest amount of compressive internal stress, i.e. –120 MPa, is given by the nickel layer on a brass mandrel electroplated at the higher current density of 5 Adm$^{-2}$ (sample St 57; see Figure 3 and Table 1).

The scatter band at the same measuring point of sample St 56 is between -50 to -100 MPa. The variation in sample St 56 could originate from a particular orientation of the nickel grains. Therefore the results may deviate with the variation in the angle of measurement.

Releasing the nickel layer from the brass mandrel, results in a decrease of the internal compressive stress. This is due to the lack of reinforcement of the thin layer by the much thicker mandrel. The internal stress must therefore relieve towards zero as confirmed by the measurements. Thus, releasing of any compressively-stressed thin electroform should result in stretching of the nickel.

In contradistinction, the results obtained by the deposit stress analyzer show another trend (Table 1). Here the measured compressive stress is higher in sample St 56 (- 21 MPa) and nearly zero (+ 1 MPa) in sample St 57. However, for the same equipment, Stein [8] has shown the compressive internal stress for nickel reducing to approximately zero as current density increases from 1 Adm$^{-2}$ to about 4.5 Adm$^{-2}$ when using a nickel sulphamate electrolyte. This result compares very favourably with our results for sample St 57, although the precise chemistry of Stein's nickel sulphamate electrolyte was not given.

Further experimental investigations are necessary to understand the relationships between results obtained under different plating conditions (e.g. flow of electrolyte, size and orientation of anode/cathode, substrate material and size) analysed by different stress measurement methods. An improved understanding of the stress characterisation will then allow the relationships between internal stress, stretch and electrolyte additives to be investigated in a more comprehensive manner.

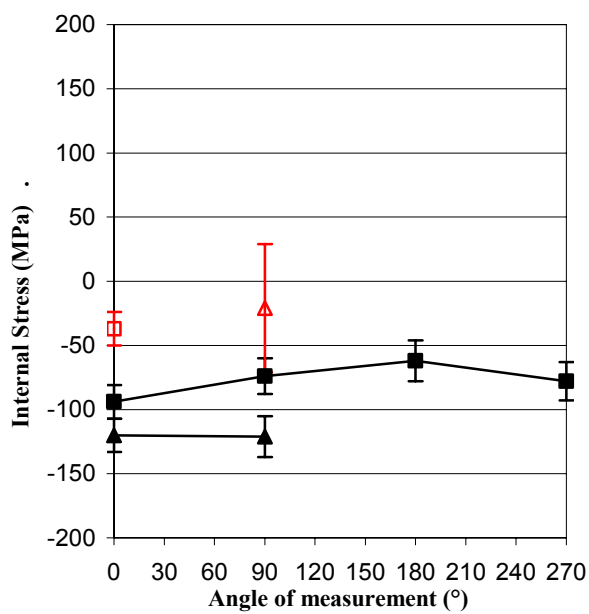

Figure 3. Stress measured by XRD for St 56 on a brass mandrel (■) and free-standing (□) and St 57 on a brass mandrel (▲) and free-standing (Δ).

### 4.2 Relationship of internal stress to stretch

The relationship between stress and stretch (measured in parts per million) is illustrated in Figure 4. The internal stress value has been derived from the deposit stress analyzer measurement and therefore gives a first approximation to the actual stress likely to be experienced by the electroformed part. The high correlation coefficient ($R^2 = 0.8167$) indicates a very strong correlation between internal compressive stress and the resulting stretch of an electroformed component despite the fact that the absolute value of the internal stress is in dispute.

### 4.3 Effects of additives on internal stress

Notwithstanding the difficulties of measuring stress accurately, some preliminary analysis on the general effects of additives on internal stress has been carried out [10]. In summary, additions of nickel chloride appear to *reduce* compressive stress. However, hardener additions not only increase the hardness of the deposit but in so doing also reduce grain size, increase the visual lustre and *increase* compressive stress as shown in Figure 5. Increasing the current density also increases the electrodeposit lustre. It is therefore vital in industrial PEF to measure and control not only the current density but also the electrolyte chemistry and additive concentrations.

### 5. CONCLUSIONS

Previous research has been carried out in comparing the values of internal stresses within electroformed nickel, obtained *via* a deposit stress analyzer, a spiral contractometer and an IS meter [9]. In this work, XRD results have been compared with deposit stress analyzer results.

Our preliminary investigations suggest that the results obtained from a deposit stress analyzer are self-consistent and allow a rapid evaluation of stress that can be used for electroplating control. The results obtained appear to give an average value for the coating relating to the particular geometrical arrangement within the electroplating cell that can be related to the perpendicular mandrel onto which a part is plated.

XRD is an extremely accurate measurement over such a very small area that, again, the stress value may not be typical of the overall stress across a much larger component. Indeed, the analysis may be so localized, and relevant only to a maximum depth of 3.5μm, that the results may not reflect the overall stress even in an electroformed micro-product.

The XRD method appears very sensitive to grain orientation that may be influenced by the mandrel crystallographic orientation and/or the current density. Even stresses induced by pulling the electroform from the mandrel may be detected. XRD also has particular value when the electroform thickness is extremely small – a condition that results in higher errors for a deposit stress analyzer [9].

However, XRD clearly confirms that on removal from a mandrel, the compressive internal stress is relieved and consequently becomes apparent in component stretch.

### 6. ACKNOWLEDGEMENTS

Prof. Allen, Prof. Saumer and Mr. Duclos wish to acknowledge the cooperation of Tecan Ltd, its staff and its Chairman and Owner, Mr. Paul Cane, and to express their thanks for access to the industrial manufacturing facilities and an insight into the complex world of commercial PEF.

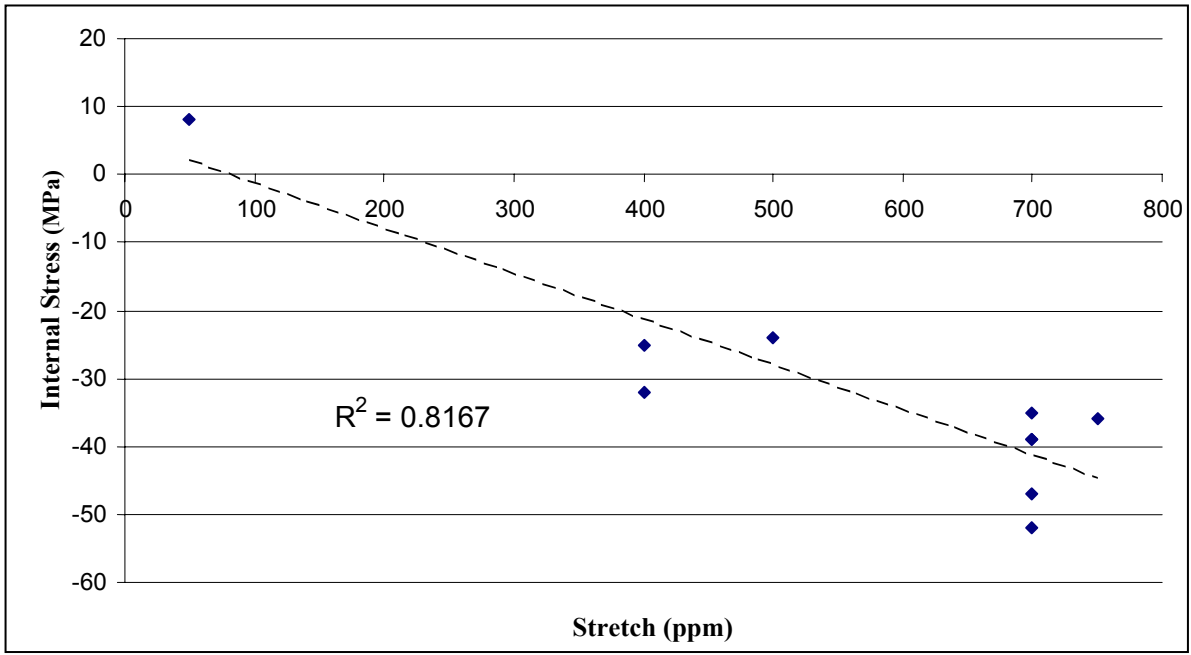

Figure 4 Graph showing direct relationship between internal stress and stretch

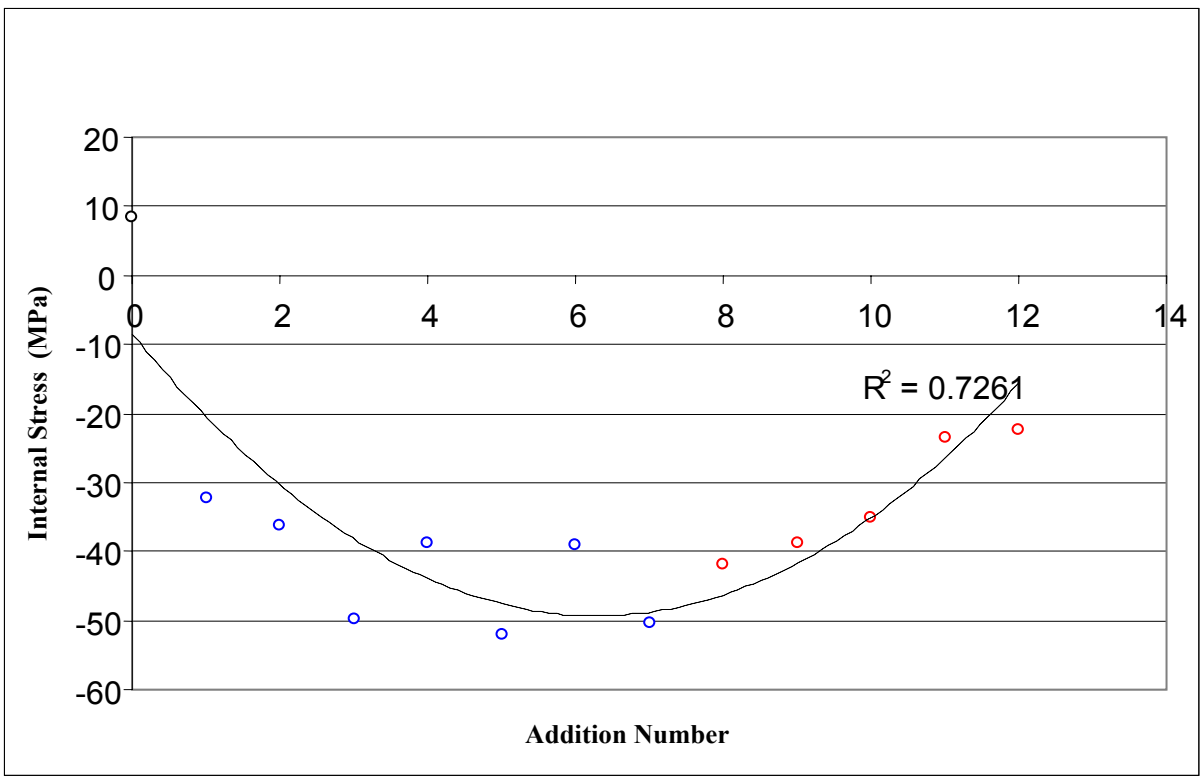

Figure 5. Graph showing effect of additions (1-7) of hardener and additions (8-12) of nickel chloride.

# 7. REFERENCES

[1] L. Uriarte, A. Ivanov, H. Oosterling, L. Staemmler, P.T. Tang and D. Allen, "A comparison between microfabrication technologies for metal tooling", Proceedings of the International Conference on Multi-Material Micro Manufacture (4M), Karlsruhe, Germany, pp. 351-354, 29[th] June – 1[st] July 2005.

[2] D.M. Allen, H.J.A. Almond, N. Hollinshead and G. Kayode Ake, "Comparisons of various physical characteristics of electroformed nickels and nickel-cobalt alloys as an aid for the selection of micro-mould insert materials", Design, Test, Integration and Packaging of MEMS/MOEMS Proceedings, Montreux, Switzerland, pp. 360-362, 1[st] -3[rd] June, 2005.

[3] E. Macherauch, H. Wohlfahrt, U. Wolfstieg. „Zur Zweckmäßigen Definition von Eigenspannungen", Härterei Tech. Mitt. 28, 1973, pp. 201-211.

[4] V. Hauk, Structural and Residual Stress Analysis by Nondestructive Methods, Elsevier, Amsterdam, 1997.

[5] B. Stein, A practical guide to understanding, measuring and controlling stress in electroformed metals, AESF Electroforming Symposium, Las Vegas, NV, USA, March 27[th]-29[th], 1996.

[6] http://www.specialtytest.com accessed on 29[th] January, 2006

[7] E. Macherauch und P. Müller, Das $\sin^2\psi$-Verfahren der röntgenographischen Spannungsmessung, Z. angew. Physik, 13, 1961, pp. 305-312.

[8] B. Stein, Fast and accurate deposit internal stress determination, Session N- Quality in Surface Finishing, Sur/Fin'2000, Navy Pier, Chicago, IL, USA accessed as http://www.specialtytest.com/ST2000.htm on 9th February, 2006.

[9] G. Richardson and B. Stein, Comparative study of three internal stress measurement methods, AESF Electroforming Symposium, San Diego, CA, USA, October 1[st] -3[rd], 1997 accessed through reference [6].

[10] N. Duclos, The effects of additives on the physical properties of electroformed nickel, MSc thesis, Cranfield University, Bedford, UK, September 2005.